\documentclass[aps,prx,twocolumn,superscriptaddress,longbibliography]{revtex4-1}
\usepackage{mathrsfs}
\usepackage{tabularx}
\usepackage{slashed}
\usepackage{amsmath}
\usepackage{amsfonts}
\usepackage{graphicx,xcolor}
\usepackage{times}
\usepackage[caption=false]{subfig}
\usepackage[colorlinks,linkcolor=blue,anchorcolor=red,citecolor=blue]{hyperref}
\usepackage{amsthm}
\usepackage{amssymb, amsbsy, latexsym, dsfont, array, layout}
\usepackage{braket}

\newcommand{\ketbrad}[1]{\left|{#1}\rangle\!\langle{#1}\right|}
\newcommand{\ketbra}[2]{\left|{#1}\rangle\!\langle{#2}\right|}

\begin{document}

\title{Decoherence Resilience of the Non-Hermitian Skin Effect}

\author{Kunkun Wang}
\affiliation{School of Physics, Anhui University, Hefei 230601, China}
\affiliation{School of Physics, Southeast University, Nanjing 211189, China}
\author{Lei Xiao}
\affiliation{School of Physics, Southeast University, Nanjing 211189, China}
\author{Stefano Longhi}\email{stefano.longhi@polimi.it}
\affiliation{Dipartimento di Fisica, Politecnico di Milano, Piazza Leonardo da Vinci 32, I-20133 Milano, Italy}
\affiliation{IFISC (UIB-CSIC), Instituto de Fisica Interdisciplinary Sistemas Complejos, E-07122 Palma de Mallorca, Spain}
\author{Peng Xue}\email{gnep.eux@gmail.com}
\affiliation{School of Physics, Southeast University, Nanjing 211189, China}

\begin{abstract}
\textbf{
Decoherence and dissipation, arising from unavoidable interactions with the environment, can exert a dual influence on transport in physical systems, suppressing coherent propagation while inducing diffusion and mitigating localization in disordered systems. Non-Hermitian physics reveals a qualitatively different scenario, in which structured dissipation can induce directional bulk-to-boundary transport, known as the non-Hermitian skin effect (NHSE), that remains robust against disorder. Whether such transport can persist, be enhanced or hindered under decoherence, remains a largely open question. Here we experimentally address this question using photonic quantum walks with two tunable prototypical decoherence channels, dephasing and amplitude damping. Under dephasing, the NHSE survives up to the fully incoherent regime and is observed to even be enhanced by dephasing, yielding drift velocities that exceed those of coherent dynamics. By contrast, amplitude damping shows a pronounced order dependence: applied before the non-Hermitian loss operator, it suppresses and ultimately eliminates the NHSE in the fully incoherent limit; applied afterward, the NHSE persists and can be enhanced at sufficiently large loss strengths. Our work bridges quantum and classical non-Hermitian dynamics, demonstrates the resilience of the NHSE to decoherence, and opens avenues for harnessing decoherence to enhance directional transport in noisy, nonequilibrium systems.
}
\end{abstract}

\maketitle

In realistic physical systems, unavoidable interactions with the environment lead to noise, dissipation, and decoherence, which can exert multifaceted influences on transport. While decoherence is often viewed as detrimental to coherent dynamics \cite{Z03}, it can also enhance transport in disordered systems by mitigating localization effects~\cite{PlenioHuelga2008Dephasing,Caruso2009NoiseAssisted,Viciani2015NoiseAssisted}. Recent developments in non-Hermitian physics challenge and further enriches this common scenario. Tailored dissipation has recently emerged as a powerful tool for engineering non-Hermitian dynamics with properties unattainable in Hermitian settings~\cite{B07,AGU20,Longhi2017PTphotonics,Midya2018NonHermitian,BBK21,HZW23,XWQ25,HO25}.
By allowing complex eigenvalues and relaxing the orthogonality of eigenstates, non-Hermitian systems give rise to striking phenomena, including unconventional spectral and dynamical responses~\cite{PCC20,XLW24,SHW24,LZW25,XCH25}, enhanced sensitivity to perturbations~\cite{COZ17,KCE22,XCL24}, and non-Hermitian topological phenomena~\cite{GAK18,KSU19,WDY21,SWC23,LZJ25}.

Among these phenomena, the non-Hermitian skin effect (NHSE)~\cite{YW18,KEB18,NKK20,LLM20,ZYF22,WWM22,WSW24,Lin2023TopologicalNHSE,L25,WFC25} has attracted particular attention. In Hermitian systems, introducing a boundary or interface does not significantly affect the bulk eigenmode spectrum. By contrast, in certain non-Hermitian systems, such as those with tailored non-reciprocal couplings, open boundaries or interfaces can drive all bulk modes to localize at the edge. This effect fundamentally alters the relation between bulk spectra and boundary responses, challenging the conventional bulk-boundary correspondence central to topological physics~\cite{YW18,KEB18}. Recent theoretical and experimental advances have shown that the NHSE is a universal phenomenon \cite{ZYF22} and far more than a spectral anomaly. By collectively biasing bulk states toward a boundary, it naturally induces directional transport of waves, energy, and information~\cite{LLW22,LXD22,LWW24,ZWH25}, which is robust against disorder in the system \cite{Longhi2015}. This feature enables precise control of transport~\cite{LLG20,XLW24}, wave amplification or suppression~\cite{GGX22,ZWL24}, positioning the NHSE as a versatile mechanism for directional state transfer~\cite{JZZ24}, enhanced sensing~\cite{MC20,ZZC25}, quantum reservoir computing~\cite{Sannia2025} and programmable flow~\cite{LSW24} across diverse platforms.

Despite extensive studies of the NHSE under fully or partially coherent conditions~\cite{XDW20,HNH21,KYL25,Sannia2025}, its behavior in regimes bridging the quantum-to-classical transition and encompassing different types of decoherence remains largely unexplored~\cite{L24}.
Decoherence, which naturally gives rise to incoherence, is typically expected to suppress dynamical propagation, driving the transition from quantum superdiffusion to classical diffusion~\cite{Z03,BCA03,SCP11}. Under such incoherent dynamics, it has been suggested that the NHSE is no longer a universal phenomenon~\cite{ZYF22} but can manifest differently depending on whether the system is globally reciprocal or not~\cite{L24}. Whether the highly directional bulk-to-boundary transport characteristic of the NHSE can survive, adapt to, or even be enhanced by different forms of decoherence thus remains a fundamental open question.
In addition, incoherent hopping dynamics are widespread in complex physical, chemical, and biological systems far from equilibrium~\cite{PlenioHuelga2008Dephasing,Caruso2009NoiseAssisted,Viciani2015NoiseAssisted,Engel2007,Lambert2012QuantumBiology,Alvarez2024QuantumBiology}, underscoring the broad relevance of this question for open quantum, classical, and hybrid platforms.

\begin{figure*}
\centering
\includegraphics[width=0.8\textwidth]{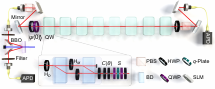}
\caption{
{\bf Experimental setup.}
Photon pairs are generated via a $\beta$-barium-borate (BBO) nonlinear crystal and initialized in a horizontally polarized Hermite-Gaussian spatial mode. The initial coin (polarization) state is prepared using a half-wave plate (HWP) and a quarter-wave plate (QWP) placed after the first polarizing beam splitter (PBS). Each step of the quantum walk (QW) is implemented by two HWPs realizing the coin operation, followed by a $q$-plate combined with two QWPs and a HWP to implement the conditional shift operator. Both the loss operator and the decoherence channels are implemented using combinations of two beam displacers (BDs) and wave plates. The walker position, encoded in orbital angular momentum, is subsequently measured using a spatial light modulator (SLM) and avalanche photodiodes (APDs) coupled via single-mode fibers.
}
\label{fig:1-setup}
\end{figure*}

Here we demonstrate the NHSE in photonic quantum walks (QWs) with tunable dephasing and amplitude damping. By engineering photon loss and decoherence via polarization control, our platform enables a continuous interpolation between coherent quantum and classical stochastic dynamics, thereby providing direct experimental access to the crossover from fully coherent to fully incoherent transport. This setting allows us to address the open question of how the NHSE behaves under decoherence and to systematically investigate its survival, deformation, and potential enhancement as coherence is reduced.

Remarkably, we show that the NHSE persists under dephasing, and can even be amplified by it. The amplification reaches its maximum in the fully incoherent limit---a behavior that stands in stark contrast to the conventional ``Goldilocks effect,'' where transport efficiency typically exhibits a non-monotonic dependence on noise and displays an optimal intermediate regime~\cite{Viciani2015NoiseAssisted,Longhi2026Lifshitz}.
In contrast, amplitude damping exhibits a pronounced order dependence: applied before the non-Hermitian operator, it suppresses and eventually eliminates the NHSE, whereas applied afterward, the NHSE survives and can even be amplified at large loss strengths. These results highlight the robustness and decoherence-assisted tunability of the NHSE, establishing a unified framework for systematically studying and engineering non-reciprocal flow in photonic systems, active-matter systems~\cite{SYK25}, and other far-from-equilibrium classical or quantum systems.

\begin{figure*}
	\centering
	\includegraphics[width=\textwidth]{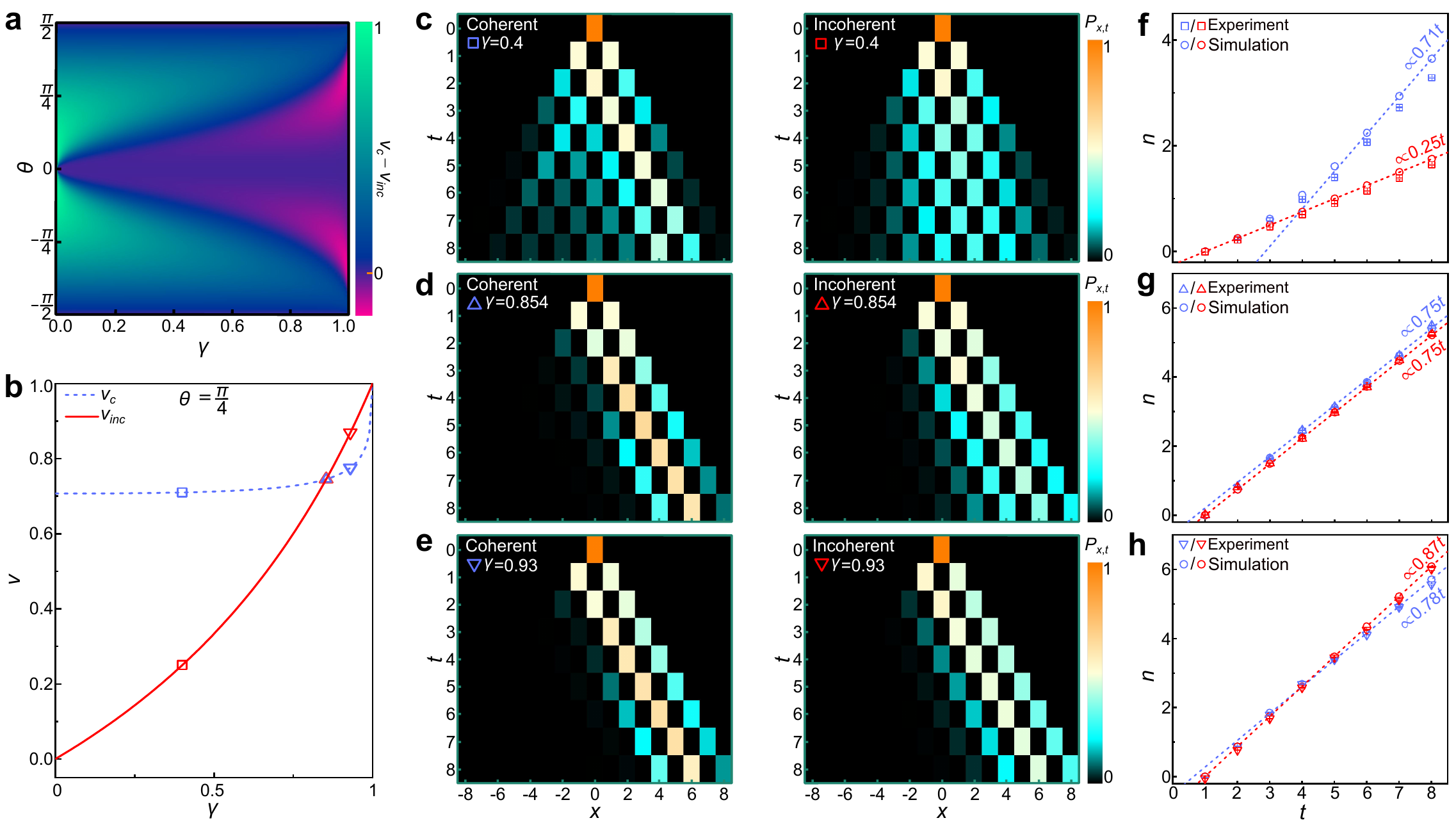}
	\caption{
{\bf Observation of the non-Hermitian skin effect under dephasing.}
{\bf a} Difference in drift velocity between coherent and fully incoherent dynamics as a function of the non-Hermitian loss strength $\gamma$ and the coin-operator parameter $\theta$.
{\bf b} Drift velocity as a function of $\gamma$ at the fixed coin parameter $\theta=\pi/4$. The blue dashed line corresponds to the coherent dynamics, while the red solid line represents fully incoherent dynamics. Symbols indicate selected experimental parameters: $\gamma=0.4$ (squares), $\gamma=0.854$ (upward triangles), and $\gamma=0.93$ (downward triangles).
{\bf c-e} Experimentally measured probability distributions during the eight-step quantum walk (QW) for $\gamma=0.4$, $0.854$, and $0.93$, respectively.
{\bf f-h} Time evolution of the center of mass $n(t)$ corresponding to {\bf c-e}. Numerical simulations are shown as open circles, while experimental data are indicated by symbols with different shapes. Blue and red symbols correspond to coherent and fully incoherent dynamics, respectively. Colored dashed lines show the theoretical fitted drift, with slopes given by the analytically calculated drift velocities from Eqs.~(\ref{eq:depC}) and~(\ref{eq:depIC}). Error bars indicate statistical uncertainties from photon-number counting.
}
\label{fig:2-dephasing}
\end{figure*}

\begin{figure}
	\centering
	\includegraphics[width=0.45\textwidth]{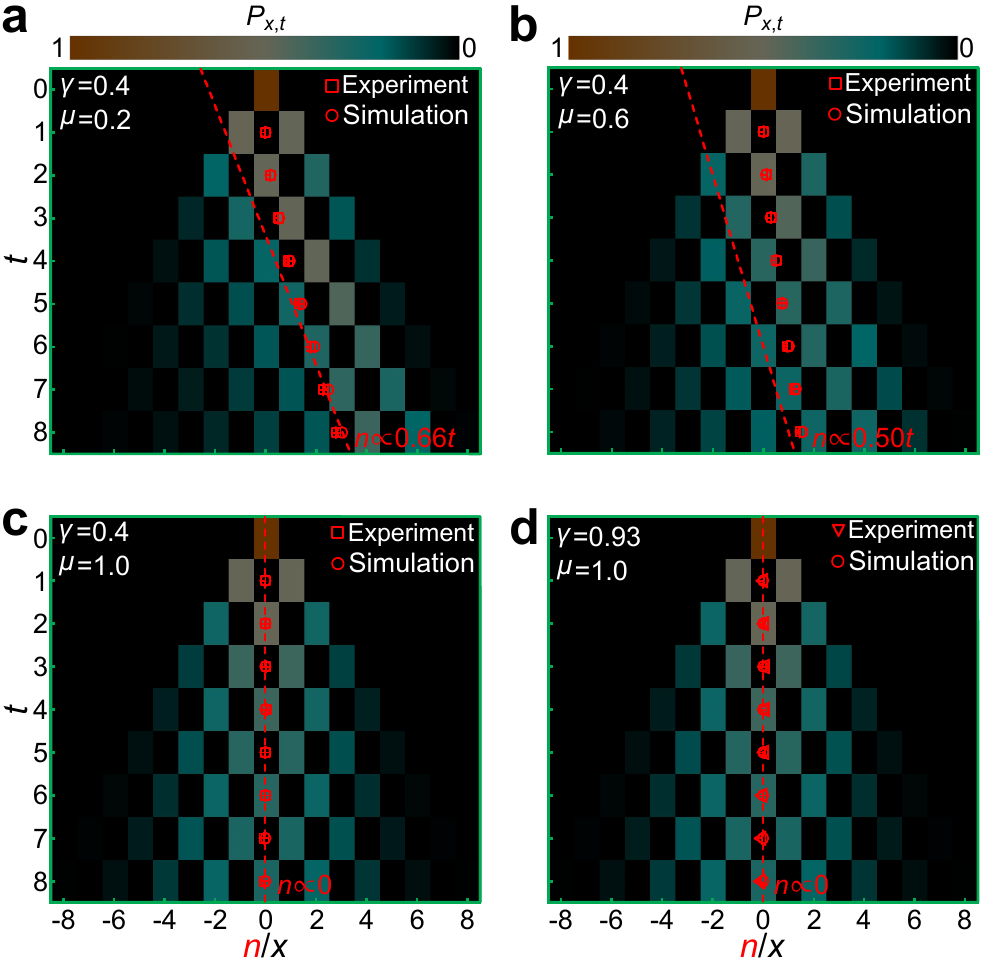}
	\caption{
{\bf Observation of the non-Hermitian skin effect under amplitude damping.}
{\bf a-c} Experimentally measured probability distributions of the eight-step QW at a fixed loss strength $\gamma = 0.4$, for increasing damping strengths $\mu = 0.2$, $0.6$, and $1.0$.
{\bf d} Probability distribution for $\gamma = 0.93$ with full damping ($\mu = 1.0$), corresponding to fully incoherent dynamics. Open red squares indicate the center of mass extracted from the measured probability distributions, while open red circles denote the corresponding numerical simulations. Red dashed lines are linear fits to the long-time center-of-mass evolution, with slopes corresponding to the drift velocities. Error bars indicate statistical uncertainties from photon-number counting.
}
\label{fig:3-damping}
\end{figure}

\begin{figure*}
	\centering
	\includegraphics[width=\textwidth]{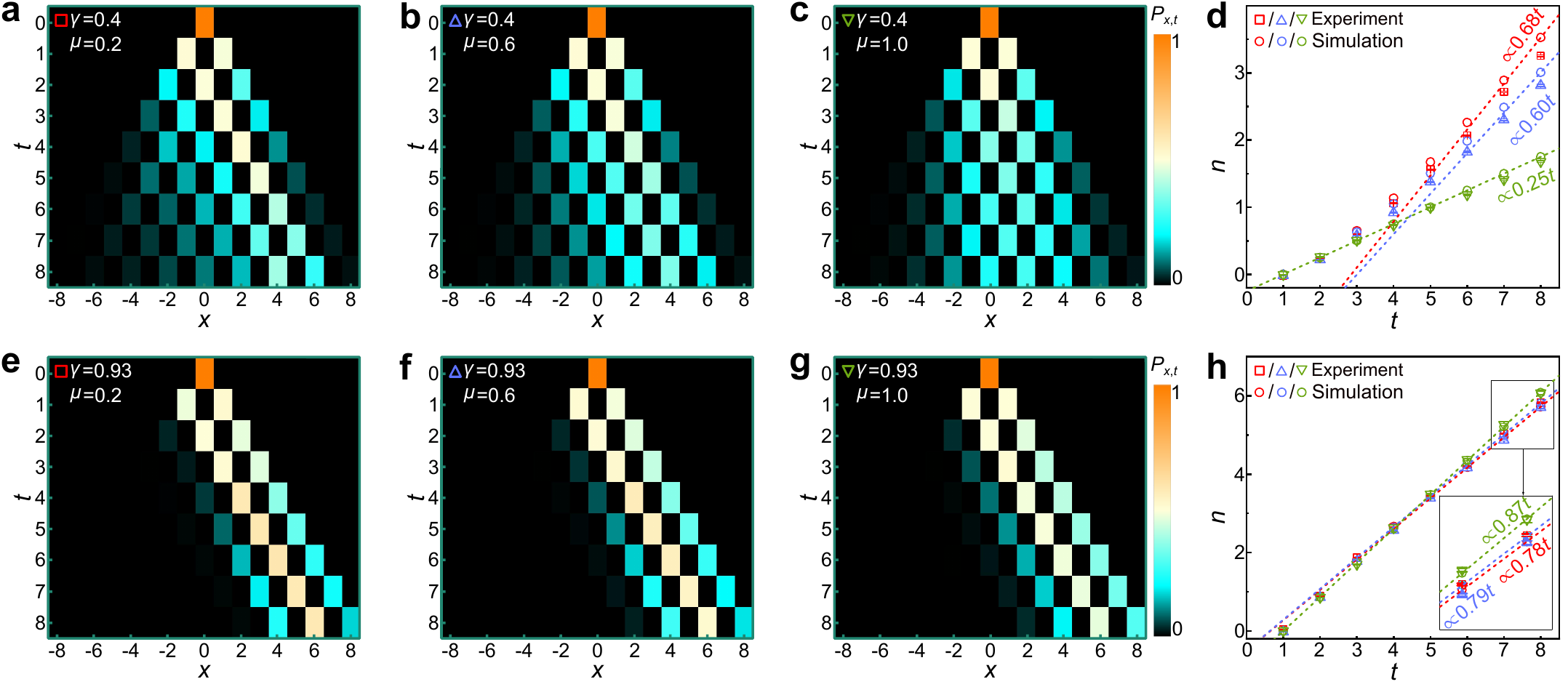}
	\caption{
{\bf Persistence of the non-Hermitian skin effect under reversed amplitude damping ordering.}
Experimentally measured probability distributions during the eight-step QW, with amplitude damping applied after the loss operator. For a fixed loss strength $\gamma=0.4$, the damping strength increases from $\mu=0.2$ in {\bf a} to $\mu=0.6$ in {\bf b} and reaching the fully incoherent limit of $\mu=1.0$ in {\bf c}. {\bf d} Time evolution of the center of mass $n(t)$ for the same parameter sets as in {\bf a-c}. Numerical simulations are shown as open circles, while experimental data are indicated by symbols of different shapes. Colored dashed lines represent linear fits to the long-time center-of-mass evolution, with slopes corresponding to the drift velocities. {\bf e-g} Experimentally measured probability distributions for a larger loss strength $\gamma = 0.93$ under varying damping strengths. {\bf h} Corresponding center-of-mass evolution for {\bf e-g}. Error bars indicate statistical uncertainties from photon-number counting.
	}
	\label{fig:4-redamping}
\end{figure*}

{\bf Results}

{\bf Implementation of QWs with decoherence.}
QWs provide a versatile platform for simulating lattice transport in diverse physical settings~\cite{XDW20,XLW24,XWQ25}. In each step, a coin operator acts on a two-level internal (coin) state, spanned by $\ket{0}$ and $\ket{1}$, followed by a conditional shift that moves the walker between discrete lattice sites $\ket{x}$, thereby generating controlled quantum interference. This produces ballistic spreading, with quantum walkers diffusing quadratically faster than their classical counterparts~\cite{Z03}. Moreover, by incorporating mode-selective loss operators, QWs can naturally simulate non-Hermitian lattices and access phenomena such as the NHSE~\cite{XDW20}.

Under this framework, the single-step evolution of the QW is governed by the operator $U=SC(\theta)M$. The mode-selective loss operator $M=\sum_x\ketbrad{x}\otimes (\ketbra{0}{0}+\sqrt{1-\gamma}\ketbra{1}{1})$ implements coin-dependent loss, where $\gamma$ sets the loss strength. The coin operator $C=\sum_x\ketbrad{x}\otimes e^{-i\theta\sigma_y}$ rotates the internal state by an angle $\theta$. Following, the conditional shift operator $S=\sum_x\ketbra{x+1}{x}\otimes \ketbrad{0}+\ketbra{x-1}{x}\otimes \ketbrad{1}$ moves the walker right or left depending on its coin state. Without loss of generality, we take the initial state $\rho_0=\ketbrad{\psi_i}$ with $\ket{\psi_i}=\ket{0}\otimes(\ket{0}+i\ket{1})/\sqrt{2}$. The system then evolves for $t$ steps as $\rho_t=U^t\rho_{0}U^{\dagger t}$.

Decoherence bridges the quantum and classical walks, providing a tunable pathway from coherent quantum evolution to classical stochastic dynamics~\cite{SCP11}. Here we focus on two ubiquitous forms of decoherence, dephasing and amplitude damping, whose combined action can effectively capture a wide variety of harmful noise processes~\cite{MFC20,MFC21}. Under dephasing, the state at each step evolves as $\rho_{t+1}=(1-\eta)U\rho_{t}U^{\dagger}+\eta\sum_iUK_i\rho_{t}K_i^{\dagger}U^{\dagger}$,
with the Kraus operators $K_i=\sum_x\ketbrad{x}\otimes\ketbrad{i}$ and $i\in\{0,1\}$~\cite{BFL10}. The parameter $0\leq\eta\leq1$ quantifies the dephasing strength, with $\eta=0$ and $\eta=1$ corresponding to fully coherent and incoherent QWs, respectively. For amplitude damping, the evolution is
$\rho_{t+1}=\sum_iU\mathcal{K}_i\rho_{t}\mathcal{K}_i^{\dagger}U^{\dagger}$, with $\mathcal{K}_0=\sum_x\ketbrad{x}\otimes(\ketbra{0}{0}+\sqrt{1-\mu}\ketbra{1}{1})$ and $\mathcal{K}_1=\sqrt{\mu}\sum_x\ketbrad{x}\otimes\ketbra{0}{1}$. Here $\mu$ ($0\leq\mu\leq1$) defines the amplitude-damping strength.

Experimentally, we encode the coin in the photon polarization, with $\ket{H}\rightarrow\ket{0}$ and $\ket{V}\rightarrow\ket{1}$, while the walker position is encoded in orbital angular momentum (OAM) states, $\ket{x}\rightarrow\ket{l}$~\cite{CMQ15,WCZ18}. As shown in Fig.~\ref{fig:1-setup}, heralded single photons are coupled into a single-mode fiber to prepare the transverse electromagnetic mode TEM$_{00}$, thereby initializing the walker in the OAM state $l=0$. Photons are initially prepared in horizontal polarization using a polarizing beam splitter (PBS). The coin state $(\ket{H}+i\ket{V})/\sqrt{2}$ is then created by passing the photons through a half-wave plate (HWP) followed by a quarter-wave plate (QWP). Note that a spatial light modulator (SLM) placed after the PBS can be used to prepare photons in different superpositions of OAM states, if desired.

Experimentally, dephasing is applied to the coin degree of freedom via probabilistic polarization operations, where the polarization is either rotated by a HWP set to $0^\circ$ or left unchanged.  The relative probabilities of these two operations are controlled through photon collection times~\cite{MFC21,CRV11,FLZ25}, allowing the dephasing strength to be tuned continuously from fully coherent to fully incoherent dynamics. The mode-selective loss operator $M$ is implemented using a pair of beam displacers (BDs) and wave plates. By adjusting the setting angle of H$_M$, the loss strength can also be continuously tuned. The unitary coin and shift operators are subsequently realized using two HWPs and a $q$-plate sandwiched between appropriately oriented wave plates~\cite{WCZ18}. After a fixed number of steps, the walker distribution is reconstructed by resolving both the coin and position degrees of freedom. The coin state is analyzed via polarization analysis (a HWP and a PBS), while the position information is obtained through OAM mode-selective detection using a SLM and single-mode fiber coupling. Amplitude damping is realized using the same optical elements employed for the loss operator. By tuning the wave-plate settings, different composite operations $M\mathcal{K}_i$ are implemented. Each operation is sampled with equal probability by allocating identical photon collection times (see Methods for more detail).

{\bf Resilience and enhancement of NHSE under dephasing.}
In QWs exhibiting the NHSE, the interplay of the mode-selective loss operator ($M$) with the shift ($S$) and coin ($C$) operators generates a robust directional transport of the wave packet.
Unlike conventional ballistic or diffusive spreading, this transport gives rise to a steady, initial-state–independent drift velocity across position space~\cite{L24}. To quantitatively capture this directional transport, we characterize the walker dynamics by the center of mass of the wave packet, defined as $n(t) = \sum_x x P(x,t)$, where $P(x,t)$ denotes the probability of detecting the walker at position $x$ during the $t$-th step~\cite{XLW24}.

Under coherent evolution, the NHSE manifests as an approximately linear growth of $n(t)$ at large step numbers, $n(t) \sim v_c(\gamma, \theta) t$~\cite{L24}. Here the coherent drift velocity $v_c$ is determined by the dominant eigenmodes of the effective Hamiltonian with the largest imaginary eigenvalues, satisfying (see Methods)
\begin{equation}
v_c(\gamma, \theta) \sim \pm \frac{(1+\sqrt{1-\gamma}) \cos(\theta)}{\sqrt{4\sqrt{1-\gamma} + (1-\sqrt{1-\gamma})^2 \cos^2(\theta)}}.
\label{eq:depC}
\end{equation}
Here the sign defines the transport direction. Under fully incoherent dynamics induced by maximal dephasing ($\eta=1$), the drift velocity is given by
\begin{equation}
v_{inc}(\gamma, \theta) \sim \pm \frac{\gamma \cos^2(\theta)}{\sqrt{4(1-\gamma) \sin^4(\theta) + \gamma^2 \cos^4(\theta)}}.
\label{eq:depIC}
\end{equation}

As shown in Fig.~\ref{fig:2-dephasing}a, we compare the drift velocities for coherent and fully incoherent QWs. Except at $\theta = 0$ or $\pm \pi/2$, where $v_c = v_{inc} = 0$ or $1$, dephasing generally suppresses the NHSE at weak loss strengths, resulting in $v_c > v_{inc}$, consistent with the conventional view that dephasing inhibits transport. Remarkably, this trend reverses as the loss strength $\gamma$ increases: the dephasing-induced suppression progressively weakens and is eventually overturned, such that dephasing enhances the transport and yields $v_c < v_{inc}$. This behavior is further illustrated in Fig.~\ref{fig:2-dephasing}b, which shows the drift velocity as a function of $\gamma$ for a fixed coin parameter $\theta = \pi/4$. For large loss strengths $\gamma$, the drift velocity increases monotonically with the dephasing rate $\eta$, reaching its largest value at $\eta = 1$~\cite{suppl}. This behavior contrast with the usual scenario where transport efficiency is optimized at an intermediate dephasing rate---the ``Goldilocks effect''---since strong dephasing typically freezes the dynamics via the quantum Zeno effect ~\cite{Viciani2015NoiseAssisted,Longhi2026Lifshitz}.
Notably, the drift velocity under fully incoherent dynamics remains finite for $0 < \gamma \leq 1$, demonstrating that the NHSE persists even in the absence of coherence.

To experimentally corroborate the above results, we probe both coherent and incoherent QW dynamics under dephasing for three representative loss strengths, $\gamma = 0.4$, $0.854$, and $0.93$ with a fixed coin parameter $\theta = \pi/4$. The corresponding measured probability distributions over eight steps are shown in Figs.~\ref{fig:2-dephasing}c-e. In all cases, the distributions exhibit a clear directional transport, confirming the persistence of the NHSE. Moreover, as $\gamma$ increases, the distributions become increasingly asymmetric and accumulate closer to the transport direction, in quantitative agreement with the enhancement of the drift velocity observed in Fig.~\ref{fig:2-dephasing}b.

From these distributions, we extract the corresponding center of mass and compare it with numerical simulations. As shown in Figs.~\ref{fig:2-dephasing}f-h, excellent agreement is observed for both coherent and fully incoherent dynamics, validating our experimental implementation. Moreover, during the later steps of the eight-step QWs, the center-of-mass evolution follows an approximately linear trend, consistent with the theoretical drift velocities given by Eqs.~(\ref{eq:depC}) and~(\ref{eq:depIC}). Specifically, at $\gamma = 0.4$, the incoherent dynamics displays a reduced slope compared to the coherent case, reflecting dephasing-induced suppression of the NHSE. At $\gamma = 0.854$, the coherent and incoherent drift velocities become nearly identical, marking a crossover point. Once $\gamma$ exceeds this value, as exemplified by $\gamma = 0.93$, the incoherent dynamics exhibits a larger drift velocity than the coherent one, demonstrating a transition from dephasing-induced suppression to dephasing-enhanced transport.
This behavior reflects an interplay between dephasing and non-Hermiticity. When non-Hermiticity is weak, coherent spreading responds strongly to weak non-reciprocity, so dephasing suppresses the NHSE-induced transport. When non-Hermiticity becomes sufficiently strong, however, dephasing enhances transport by enabling unidirectional hopping to accumulate and dominate in the incoherent regime~\cite{L24}.

{\bf Order dependence of the NHSE under amplitude damping.}
Dephasing leads to a loss of coherence without energy exchange, while amplitude damping induces irreversible population loss and thus constitutes a qualitatively distinct noise channel~\cite{MFC20,MFC21}. Under full damping ($\mu=1$), all coin states are reset to $\ket{H}$, which renders the coin-selective loss operator $M$ ineffective and eliminates the NHSE. Importantly, this behavior is order dependent: applying damping after the loss operator restores the incoherent NHSE dynamics. By contrast, reversing the order of dephasing does not alter the NHSE dynamics, as both the biased probability distributions and the associated drift velocities remain essentially unchanged~\cite{suppl}. To demonstrate this interplay, we systematically investigate amplitude damping in QWs with tunable non-Hermiticity, revealing how its strength and ordering govern the directional transport of the NHSE.

Focusing first on the effect of damping strength, we consider amplitude damping applied before the non-unitary loss operator at a fixed loss strength $\gamma = 0.4$. Figures~\ref{fig:3-damping}a-c show the measured probability distributions of an eight-step QW as the damping strength $\mu$ increases from $0.2$ to $0.6$ and finally to the fully incoherent limit ($\mu=1.0$). As the damping strength increases, the directional asymmetry and accumulation toward the biased direction gradually diminish, indicating suppression of the NHSE. This suppression is manifested as a slowing of the center-of-mass evolution, leading to a drift velocity that decreases monotonically and eventually vanishes in the fully incoherent limit. Moreover, in this regime, the walker distribution converges to a classical Gaussian profile, indicating a complete disappearance of the NHSE accompanied by a transition from quantum coherent transport to classical diffusion. Importantly, this complete suppression by amplitude damping persists even deep in the large-loss regime. As shown in Fig.~\ref{fig:3-damping}d, for $\gamma = 0.93$, the QW dynamics under fully incoherent damping ($\mu = 1$) loses its directional transport and converges to a classical Gaussian distribution, confirming that the elimination of the NHSE is independent of the loss strength.

While amplitude damping can strongly suppress or even eliminate the NHSE, its effect also depends on the order in which the damping is applied. Specifically, when the amplitude damping is applied after the loss operator (i.e., reversing the previous damping ordering), the evolution of the density matrix at each step is described by $\rho_{t+1} = \sum_i S C \mathcal{K}_i M \rho_t (S C \mathcal{K}_i M)^\dagger$. Under fully incoherent dynamics induced by maximal damping ($\mu = 1$), the corresponding drift velocity is given by
\begin{equation}
\tilde{v}_{inc}(\gamma, \theta) \sim \pm \frac{\cos^2(\theta)-(1-\gamma) \sin^2(\theta)}{1-\gamma \sin^2(\theta)}.
\end{equation}
The finite drift velocity clearly demonstrates the persistence of the NHSE in fully incoherent dynamics under this damping configuration.

Experimentally, the reversed composite operations $\mathcal{K}_i M$ are realized using the optical elements for the loss operator $M$. By tuning the wave-plate settings and collecting photons for the same duration under each operation, all $\mathcal{K}_i M$ are effectively sampled with equal probability, thereby implementing the reversed ordering of amplitude damping. We first consider a weakly non-Hermitian regime with $\gamma = 0.4$. As shown in Fig.~\ref{fig:4-redamping}a-c, increasing the damping strength gradually reduces the spatial asymmetry of the walker distribution, indicating that amplitude damping suppresses the NHSE. However, even in the fully incoherent limit, a clear directional transport persists. This behavior is quantitatively captured by the center-of-mass evolution. As shown in Fig.~\ref{fig:4-redamping}d, the drift velocity decreases monotonically with increasing damping strength. In the fully incoherent regime, it approaches a finite value, confirming that amplitude damping applied after the loss operator suppresses, but does not eliminate the NHSE.

We next turn to a strongly non-Hermitian regime with $\gamma = 0.93$, where qualitatively different behavior emerges. As shown in Fig.~\ref{fig:4-redamping}e-g, increasing the damping strength amplifies the directional asymmetry of the probability distributions. Correspondingly, the center-of-mass evolution in Fig.~\ref{fig:4-redamping}h exhibits a monotonic increase of the drift velocity with damping strength. In this regime, amplitude damping gives rise to a clear noise-enhanced drift under fully incoherent dynamics.


{\bf Discussion.}
We report the first experimental demonstration of the NHSE across the range from coherent to fully incoherent dynamics, achieved by introducing controllable dephasing and amplitude damping in non-unitary QWs. While decoherence in conventional QWs typically suppresses ballistic spreading, driving a transition toward Gaussian diffusion~\cite{BCA03,SCP11}, we find that the directional transport induced by the NHSE not only persists but can be enhanced by dephasing. This establishes incoherent processes as an unexpected resource for stabilizing non-Hermitian transport.

The impact of amplitude damping is markedly order-dependent: when applied before the non-Hermitian operation, it suppresses or even removes the NHSE, whereas the reverse ordering preserves the effect and enables noise-enhanced drift. These contrasting behaviors demonstrate that the interplay between non-Hermiticity and decoherence is fundamentally nontrivial and highly sensitive to the structure of noise.

Taken together, our results reveal a previously unexplored robustness of the NHSE that spans the quantum-to-classical transition. They show that directional flow in non-Hermitian systems can survive -- and even benefit -- from realistic sources of noise. This opens a pathway for controlling transport in noisy, far-from-equilibrium systems and suggests new design principles for resilient photonic and quantum technologies.

~\\
{\bf Methods}

{\bf Experimental setup.}
To implement QWs with decoherence, we encode the internal coin degree of freedom in the photonic polarization and the position degree of freedom in discretized OAM modes. For dephasing acting on the coin states, the density-matrix evolution at each step can be rewritten as
$\rho_{t+1} = \left(1-\eta/2\right) U \rho_t U^\dagger + \eta U D \rho_t D^\dagger U^\dagger /2 $,
where $D=\sum_x\ketbrad{x}\otimes\sigma_z$ denotes the dephasing operator, and $\sigma_z$ is one of the Pauli matrices. Experimentally, the operator $D$ is realized by applying a HWP (H$_D$) oriented at $0^\circ$ to the polarization of the photons. The dephasing channel is thus implemented by probabilistically either leaving the polarization unchanged or applying the HWP at $0^\circ$ with probabilities $1-\eta/2$ and $\eta/2$, respectively. These probability weights are realized by adjusting the photon collection times. Specifically, by setting the collection-time ratio $t_0:t_1 = (1-\eta/2):(\eta/2)$, we realize dephasing channels with a tunable strength $\eta$~\cite{MFC21}.

After applying the loss operator $M$, photons with horizontal polarization remain unchanged, whereas vertically polarized photons experience loss with probability $\gamma$. Experimentally, this operator is implemented by first separating the two polarization components into distinct spatial modes using a BD, with vertically polarized photons directed into the upper path and horizontally polarized photons into the lower path.
A HWP, denoted as H$_M$, is inserted in the upper path and set at an angle $\arcsin\sqrt{1-\gamma}/2$, thereby introducing a controlled photon loss for the vertically polarized component. Two additional HWPs, both set to 45$^\circ$, are placed in the lower path and after the second BD, respectively, to recombine the polarization components via BD and complete the implementation of the loss operator.

For amplitude damping, the density matrix $\rho$ evolves at each time step according to
$\rho_{t+1}=U\mathcal{K}_0\rho_t\mathcal{K}_0^{\dagger}U^{\dagger}+U\mathcal{K}_1\rho_t\mathcal{K}_1^{\dagger}U^{\dagger},
\label{eq:damp}$
where $\mathcal{K}_0$ and $\mathcal{K}_1$ are the Kraus operators describing amplitude damping. By defining $U_i=U\mathcal{K}_i$ ($i=0, 1$), the evolution can be  written as $\rho_{t+1}=\sum_iU_i\rho_tU_i^{\dagger}$. Each operator $U_i$ can be further decomposed as $U_i=SC(\theta)M_i$ with $M_1=M\mathcal{K}_0=\sum_x\ketbrad{x}\otimes (\ketbra{0}{0}+\sqrt{(1-\gamma)(1-\mu)}\ketbra{1}{1})$ and $M_2=M\mathcal{K}_1=\sum_x\ketbrad{x}\otimes \sqrt{\mu}\ketbra{0}{1}$. In our experiment, the operator $M_1$ is implemented by adjusting the setting angle of the HWP (H$_M$) to $\arcsin\sqrt{(1-\gamma)(1-\mu)}/2$. The operator $M_2$ removes the horizontal-polarized component and converts vertically polarized photons into the horizontal polarization with probability $\mu$. Experimentally, this operator is realized by setting the angle of H$_M$ to $\arcsin\sqrt{\mu}/2$, while fixing the two HWPs in the lower path and after the second BD at 0$^\circ$. By assigning identical photon collection times to the two operations $M_i$, we implement a tunable amplitude-damping channel before the loss operator with damping strength $\mu$. Similarly, for the reversed ordering of amplitude damping, the evolution is described by the modified operators $\tilde{U}_i=SC(\theta)\tilde{M}_i$. In this case, we have $\tilde{M}_1=\mathcal{K}_0M=M_1$ and $\tilde{M}_2=\mathcal{K}_1M=\sum_x\ketbrad{x}\otimes \sqrt{\mu(1-\gamma)}\ketbra{0}{1}$. Experimentally, we realize the operator $\tilde{M}_2$ by setting the angle of H$_M$ to $\arcsin\sqrt{\mu(1-\gamma)}/2$.

After the loss operator, two HWPs with the setting angles of $0$ and $\theta/2$, respectively, are applied to implement the coin operator $C(\theta)$. To realize the shift operator, we use the $q$-plate, an optical element consisting of a thin liquid-crystal layer with a singular pattern of optic axes characterized by a topological charge $q$. In our experiment, we choose $q=1/2$, such that the OAM of the photon is shifted by $2q=1$ upon passing through the $q$-plate. Moreover, The OAM shift induced by the $q$-plate is polarization dependent and can be described by $S_q=\sum_x\ketbra{l+2q}{l}\otimes\ketbra{R}{L}+\ketbra{l-2q}{l}\otimes\ketbra{L}{R}$, where $\ket{L}=(\ket{H}+i\ket{V})/\sqrt{2}$ and $\ket{R}=(\ket{H}-i\ket{V})/\sqrt{2}$ denote the left- and right-circular polarization states, respectively. Accordingly, the shift operator is implemented using a QWP--$q$-plate--QWP--HWP sequence, with the wave plates oriented at 135$^\circ$, 135$^\circ$ and 0$^\circ$, respectively~\cite{WCZ18}.

{\bf Calculation of the drift velocity.}
The transport properties of QWs are commonly characterized by the center of mass $n(t)$. For non-unitary QWs exhibiting the NHSE, the center of mass exhibits asymptotically linear growth at long times, $n(t)\sim v t$, which defines the drift velocity $v$. In the absence of decoherence, the dynamics are governed by the non-unitary evolution $U=SC(\theta)M$. By transforming to momentum space, the evolution decouples into independent $k$-sectors described by an effective non-Hermitian Hamiltonian $H_k$ with $U_k=e^{-ikH_k}$.
In the long-time limit, the dynamics are dominated by the eigenmode associated with the eigenenergy $E_{\pm}$
having the largest imaginary part~\cite{L24}. Accordingly, the coherent drift velocity is given by
\begin{align}
v_c=\text{Re}\left(\frac{dE_{\pm}(k)}{dk}\right)\bigg|_{k=k^*},
\end{align}
where $k^*$ denotes the quasimomentum at which Im($E_{\pm}$) attains its maximum.

In the fully incoherent limit, coherence between coin states is completely suppressed, and the dynamics reduce to a classical Markov process. In momentum space, the evolution is governed by a transition matrix $\mathcal{U}_k=e^{\mathcal{M}_k}$, where $\mathcal{M}_k$ is the corresponding Markov transition matrix. The drift velocity is then determined by the eigenvalue $\lambda_{\pm}$ of $\mathcal{M}_k$. Analogously, the incoherent drift velocity is given by
\begin{align}
v_{inc}=-\text{Im}\left(\frac{d\lambda_{\pm}(k)}{dk}\right)\bigg|_{k=k^*},
\end{align}
where $k^*$ here denotes the quasimomentum at which Re($\lambda_{\pm}$) is maximized. In the partly coherent regime $0<\eta<1$, an analytical determination of the drift velocity becomes prohibitively cumbersome, and generally infeasible, because the dynamics intertwine different $(k, k')$ Fourier modes, {\em yet} the drift velocity can still be obtained numerically from the long-time scaling of the center-of-mass displacement,
\begin{equation}
v=\lim_{t\to\infty}\frac{\delta n(t)}{\delta t}.
\end{equation}

\bibliographystyle{naturemag}
\bibliography{your-references-here}

~\\
\noindent{\bf Acknowledgements}
This work is supported by the National Key R\&D Program of China (Grant No. 2023YFA1406701) and the National Natural Science Foundation of China (Grant Nos. 12025401, 92265209, 12474352 and 92476106).
S.L. acknowledges the Spanish Agencia Estatal de Investigacion (Grant No. MDM-2017-0711).
K.K.W. acknowledges support from the Natural Science Foundation of Anhui Province (2508085Y002). K.K.W. and L.X. acknowledge support from Beijing National Laboratory for Condensed Matter Physics (No. 2024BNLCMPKF010).

~\\
\noindent{\bf Author Contributions}
K.K.W. performed the experiments with contributions from L.X.. K.K.W. analyzed the experimental data, and wrote part of the paper. S.L. developed the theoretical aspects and revised the paper. P.X. designed the experiments, analyzed the results and wrote part of the paper.

~\\
\noindent{\bf Competing Interests}
The authors declare no competing interests.

~\\
\noindent{\bf Data availability}
The data that support the findings of this study are available from the corresponding authors upon requests.

~\\
\noindent{\bf Additional Information}\\
{\bf Supplementary information} The online version contains supplementary material available at XXX

\end{document}